\documentclass[useAMS,usenatbib]{mn2e}

\usepackage{amsfonts, amssymb}
\usepackage{graphicx, epsfig,bm}

\newcommand{\be}{\begin{equation}}
\newcommand{\ee}{\end{equation}}
\newcommand{\bea}{\begin{eqnarray}}
\newcommand{\eea}{\end{eqnarray}}

\def\fun#1#2{\lower3.6pt\vbox{\baselineskip0pt\lineskip.9pt
\ialign{$\mathsurround=0pt#1\hfill##\hfil$\crcr#2\crcr\sim\crcr}}}

\newcommand\lsim{\mathrel{\rlap{\lower4pt\hbox{\hskip1pt$\sim$}}
    \raise1pt\hbox{$<$}}}
\newcommand\gsim{\mathrel{\rlap{\lower4pt\hbox{\hskip1pt$\sim$}}
    \raise1pt\hbox{$>$}}}

\def\dslash{\not{\hbox{\kern-2pt $\partial$}}}
\def\Dslash{\not{\hbox{\kern-4pt $D$}}}
\def\Oslash{\not{\hbox{\kern-4pt $O$}}}
\def\Qslash{\not{\hbox{\kern-4pt $Q$}}}
\def\pslash{\not{\hbox{\kern-2.3pt $p$}}}
\def\kslash{\not{\hbox{\kern-2.3pt $k$}}}
\def\qslash{\not{\hbox{\kern-2.3pt $q$}}}

 \newtoks\slashfraction
 \slashfraction={.13}
 \def\slash#1{\setbox0\hbox{$ #1 $}
 \setbox0\hbox to \the\slashfraction\wd0{\hss \box0}/\box0 }

\def\ee{\end{equation}}
\def\be{\begin{equation}}

\title[Bayesian Evidence for a Cosmological constant]{Bayesian
  Evidence
for a Cosmological Constant using new High-Redshift Supernovae Data}
\author[P. Serra et al.]{Paolo Serra$^1$, Alan Heavens$^2$,
Alessandro Melchiorri$^1$\\
$^1$Dipartimento di Fisica e Sezione INFN, Universit\`a
degli Studi di Roma ``La Sapienza'', Ple Aldo Moro 5,00185, Rome,
Italy\\
$^2$SUPA, Institute for Astronomy, University of Edinburgh,
Royal Observatory, Blackford Hill, Edinburgh EH9-3HJ, UK }
\begin{document}

\maketitle

\label{firstpage}

\begin{abstract}
We carry out a Bayesian model selection analysis of different dark
energy parametrizations using the recent luminosity distance data of
high redshift supernovae from Riess et al. 2007 and from the new
ESSENCE Supernova Survey. Including complementary cosmological
datasets, we found substantial evidence ($\Delta \ln (E) \sim 1$)
against a time-varying dark energy equation of state parameter, and
against phantom dark energy models.  We find a small preference for
a standard cosmological constant over accelerating non-phantom
models where $w$ is constant, but allowed to vary in the range $-1$
to $-0.33$.
\end{abstract}


\section{Introduction}

Over the last few years, observations of luminosity distances of
Type Ia supernovae (SN-Ia) have established that the expansion of
the Universe is
accelerating(see e.g. \citet{riess98}, \citet{pelmutter99}, \citet{riess04},
\citet{astier2005}, \citet{bassett04}). This result is now well confirmed and complemented
by a large amount of independent observations such as, for example, the
angular-diameter distance vs. redshift relation measured by Baryonic
Acoustic Oscillations (BAO) experiments \citet{eisen05}, the
distortion of background images measured by weak lensing experiments
\citet{jarvis05}, the distance to the last scattering surface
measured by Cosmic Microwave Background (CMB) experiments
\citet{spergel06}, galaxy clustering (Large Scale Structure,
LSS)(see \citet{efstathiou02,tegmark06}) and, finally, the Integrated
Sachs Wolfe effect, correlating LSS with CMB (see e.g
\citet{giannantonio06}).

The recent analysis of \citet{riess2007} has further confirmed in an
impressive way these results, reporting the discovery of 21 new
SN-Ia with the Hubble Space Telescope (HST).
Together with a recalibration of previous HST-discovered SN-Ia,
the full sample of $23$ SN-Ia at $z > 1$ provides the
highest-redshift sample known. This dataset has then been analyzed
in combination with some of the aforementioned datasets providing new
 constraints on several dark energy properties
(see e.g. \citet{riess2007}, \citet{alam06}, \citet{gong06}, 
\citet{nesseris}, \citet{barger2006}).

This increasing quality and number of experimental datasets is
finally opening the possibility of falsifying cosmological theories
and of discriminating between different theories. There have been
many proposed explanations for this acceleration: the Einstein's
cosmological constant, a new fluid with negative pressure (constant
or varying with time) (see e.g. \citet{peebles}) or a modification of
general relativity (see for example \citet{dgp}). However, to date,
none of them is supported by a well-established fundamental theory.
Moreover, since we know (almost) nothing about dark energy there is
in principle no theoretical limit to the number of parameters that
one might use to characterize it.

It is therefore timely not only to constrain the parameters of a
specific dark energy model but also to establish reliable criteria
to choose between different models.

As pointed out in \citet{liddle} and \citet{liddle2},
there is an important
difference between \emph{parameter fitting} and \emph{model
selection}.  In the first case we work in the context of a single
theoretical model framework, and establish which choice of model
parameters gives the best fit to the data.  In a Bayesian
interpretation, with flat priors the most likely model parameters will
simply have the maximum likelihood and the lowest $\chi^2$ regardless
of the degrees of freedom. More complicated models (with a larger
number of free parameters) will normally produce better fits. However,
in model selection, we wish to know which model is favoured,
regardless of the values of the parameters.  In choosing models, we
should also look for simplicity, following somewhat a principle based
on Occam's Razor. Simplicity is of course in contrast with more
parameters and we need then to weight the ``need'' for
extra parameters. This can be accomplished by using the {\em Bayesian
Evidence}, defined as the probability of the model given the data, and
given by the average likelihood of a model over its prior parameter
space:
\begin{equation}
E=P(\vec{D}|H)=\int\,d\vec{\theta}\,P(\vec{D}|\vec{\theta},H)P(\vec{\theta},H),
\end{equation}
where $H$ is the model considered, $\vec{\theta}$ is the vector of
the model parameters, $\vec{D}$ is the data, $P(\vec{\theta},H)$ is
the prior and $P(\vec{D}|\vec{\theta},H)$ is the likelihood.  For
flat priors on the models, the probability of the model given the
data is proportional to the Evidence.

In general, comparing two models, the term
$P(\vec{D}|\vec{\theta},H)$ will be larger for the more complicated
one (which has more parameters or the same number of parameters but
with larger prior space), but, at the same time, it will have a
lower $P(\vec{\theta},H)d\vec{\theta}$ compared to the simpler
model. The first term indicates how well the model fits data, the
second one indicates how simple is the model. Following the equation
above we can assume that the best model will be the one with the
greatest Evidence.\footnote{We
  want to stress that the ``best'' model might not be in general the
  ``true model''; Nature decides what is true, not our personal
  aesthetic behaviour in simplicity.}

\citet{jef} provides a useful guide to discriminate
the difference between two models with $E_1$ and $E_2$:
\begin{equation}
1<\Delta\,\ln(E)<2.5\,\,  (substantial)\\
\end{equation}
\begin{equation}
2.5<\Delta\,\ln(E)<5\,\,   (strong)\\
\end{equation}
\begin{equation}
5<\Delta\,\ln(E) (decisive).\\
\end{equation}
The Bayesian Evidence can therefore help us to choose between models
because it establish a tension between the simplicity of a model and
its power of fitting data.

In this paper we aim exactly to make use of cosmological model
selection methods in order to discriminate between dark energy
models. In the next sections we therefore compute Bayesian Evidence
for a large set of models and we compare this value with the one
obtained for the concordance model: a flat universe with cold matter
and a cosmological constant. More specifically, in Section 2 we
present our analysis method and the dark energy models considered.
In Section 3 we show the results of our analysis and we derive our
conclusions in Section 4. Our paper follows the research lines
already investigated by previous papers (see e.g. \citet{liddle3},
\citet{saini}, \citet{multa} and \cite{polish}) 
which we will complement and extend by using a new
and independent algorithm for computing evidence (which we will
illustrate in Section 2.3), a larger set of dark energy models and
finally by considering more recent SN-Ia datasets.

\section{Analysis Method}

\subsection{Theoretical Framework}

We restrict our analysis to flat, Friedmann-Lemaitre
universes, with the redshift evolution of the expansion rate
given by:
\begin{equation}
H^2(z)=H^2_0\left[\Omega_m(1+z)^3+(1-\Omega_m)\frac{\rho_X(z)}
{\rho_X(0)}\right]
\end{equation}
\noindent where $\Omega_m$ is the energy density parameter in
matter, $H(z)$ is the Hubble constant and $\rho_X$ is the dark
energy density given by:
\begin{equation}
\frac{\rho_X(z)}{\rho_X(0)}=\exp\left\{\int_0^z\frac{3\left[1+w(z')\right]}{1+z'}dz'\right\}.
\end{equation}
\noindent where $w(z)$ is the equation of state parameter defined as
the ratio of pressure over density of the dark energy component
$P_X=w(z)\rho_X c^2$.

We consider different parametrizations of the dark energy equation
of state parameter $w(z)$. The simplest model is the usual flat
$\Lambda$CDM model with fixed equation of state $w=-1$ (MODEL I). We
then let $w$ vary, assuming it is small enough to lead to
acceleration. MODEL II has constant $w$ with a flat prior in the
range $-1\leq\,w\leq\,-0.33$. In MODEL III, we expand the prior
range to allow phantom dark energy models, constant $w$ with a flat
prior $-2\leq\,w\leq\,-0.33$. We also consider dynamical dark energy
models where $w$ can depend on redshift. In particular we consider a
linear dependence on scale factor $a=(1+z)^{-1}$ as \citet{Chevallier01}:
\begin{equation}
w(a)=w_0+w_a(1-a)
\end{equation}
\noindent with $-2\leq\,w_0\leq\,-0.33$ and
$-1.33\leq\,w_a\leq\,1.33$ (MODEL IV). The above model is a low
redshift approximation that may break at higher redshift. In this
respect it is useful to include a more sophisticated parametrization
that takes in to account the high redshift behaviour. We consider
two possibilities. The first one is the one proposed by Hannestad
and Mortsell (see \citet{hannestad}), where:
\begin{equation}
w(a)=w_0w_1\frac{a^q+a_{s}^{q}}{w_1a^q+w_0a_s^q}.
\end{equation}
\noindent In this model (MODEL V), the equation of state changes
from $w_0$ to $w_1$ around redshift $z_s=1-1/a_s$ with a gradient
transition given by $q$. The priors are flat within $-2\leq\,w_0\leq\,0$,
$-2\leq\,w_1\leq\,0$ and $0\leq\,q\leq\,10$.

The second one is the parametrization introduced by
\citet{ishak} where
\begin{equation}
w(z)=w_0+w_1z
\end{equation}
\noindent for $z<1$ and
\begin{equation}
w(z)=w_0+w_1
\end{equation}
\noindent for $z\geq1$ (MODEL VI). In this case we choose flat priors in $-2\leq\,w_0\leq\,-0.2$ and
$-4\leq\,w_1\leq\,2$.
In analyzing each model, the priors for the other parameters are flat within the ranges\\
$0.1\leq\,\Omega_m\leq\,0.5$\\
$56\leq\,H_0/(km\,s^{-1}\,Mpc^{-1})\leq\,72$.

\subsection{Cosmological Datasets}

The dark energy models are then compared with the data following the
approach described in \citet{wang3, liddle3}. In particular we
compare the luminosity distance at redshift $z$ of each model given
by
\begin{equation}
d_L(z)=c (1+z)\int_0^z \frac{dz'}{H(z')}
\end{equation}
\noindent with the SN-Ia luminosity distances from the latest
catalogue of \citet{riess2007}. This includes $182$ SN-Ia,
``flux-averaged'' with a $\Delta z=0.05$ binning as in
 \citet{wang} to reduce possible systematic effects from weak
lensing. We also consider the new $57$ supernovae coming
from the ESSENCE Supernova Suvey of \citet{wood} in combination
with the $38$ nearby supernovae (with $<0.023<z<0.15$) of
\citet{riess2007}.
We also consider the CMB shift parameter
 $R$ measured by the three-year
WMAP experiment, $R=1.70\pm0.03$ \citet{spergel06}, in combination
with the BAO measurement of the distance parameter at redshift
$z=0.35$, $d_V(z=0.35)=1.300\pm0.088$ Gpc (see \citet{eisen05}). The
shift parameter $R$ is defined as
\begin{equation}
R=\Omega_m^{\frac{1}{2}}\int_0^{z_{CMB}}\frac{dz'}{H(z')},
\end{equation}
\noindent where $z_{CMB}=1089$ is the redshift of recombination.
For the BAO measurement, the distance parameter is:
\begin{equation}
d_V(z_{BAO})=\left[r^2(z_{BAO})\frac{cz_{BAO}}{H(z_{BAO})}\right]^{\frac{1}{3}}
\end{equation}
where $r(z)$ is the comoving distance at redshift $z$ and
$z_{BAO}=0.35$. We decide not to use data coming from weak
gravitational lensing and galaxy clustering because they have the
largest systematics and we prefer to be conservative in our
analysis.

\subsection{A new algorithm for computing Bayesian Evidence}

Given a set of cosmological data, we evaluate the Bayesian Evidence
by integrating the likelihood distribution with a method based on a
modified version of the VEGAS algorithm. Introduced by
\citet{lepage}, VEGAS is widely used for multidimensional problems
which occur in elementary particle physics. VEGAS is an
importance-sampling algorithm, where regions where the integrand has
large absolute value are sampled with a higher density of points
than regions where it is low.  The key element of the VEGAS
algorithm is that samples are drawn from a probability distribution
which is separable in the coordinates. This reduces complexity in
two ways.  In a $d$-dimensional problem, the probability
distribution $p$ is specified by $d$ one-dimensional distributions,
rather than one $d$-dimensional distribution:
\begin{equation}
p(\vec\theta) \propto g_1(\theta_1)g_2(\theta_2)\ldots
g_d(\theta_d).
\end{equation}
Secondly, the generation of the samples is simplified by
successively drawing from each of the $d$ probability distributions.
The optimal weight function can be shown to be \citep{lepage}
\begin{equation}
g_i(\theta_i) \propto \sqrt{\int_{j\ne i} d^{d-1}\theta_j
\frac{f^2(\vec\theta)}{\prod_{j\ne i} g_j(\theta_j)}}
\end{equation}
where $f(\vec\theta)$ is the integrand to be sampled.  An initial
(e.g. random) sampling estimates $f$ roughly, from which an initial
set $g_i$ can be constructed.  Subsequent samples drawn from the
$g_i$ can then be used to refine the $g_i$.  See \citet{NR} for
further details. A problem of VEGAS is that it may not do well when
the integrand is concentrated in one-dimensional (or higher) curved
trajectories (or hypersurfaces), unless these happen to be oriented
close to the coordinate directions.

To solve this problem we have generalised the algorithm, and
calculate, for a first and preliminary sampling, the covariance
matrix of the Likelihood function. This we diagonalise and use the
eigenvectors to define new parameters. In terms of these new
parameters, the likelihood should be closer to separable.  Note that
this modified VEGAS algorithm should be particularly efficient if
the likelihood is single-peaked.  For multimodal likelihoods with
more significant peaks, it will become increasingly less efficient,
depending on the number, shape, relative height and orientation of
the peaks.

Given a cosmological dataset and a theoretical framework, the
Likelihood function will be clearly a function of $\vec{\theta}$:
\begin{equation}
L(\vec{\theta})\propto\,\exp\left[-\frac{\chi^2(\vec{\theta})}{2}\right]
\end{equation}
and the algorithm calculates
\begin{equation}
I=\int\,d\vec{\theta} L(\vec{\theta}).
\end{equation}
We always consider flat priors in our analysis; this implies that
the priors will be constant so the Evidence will
be given by:
\begin{equation}
E=P\cdot\,I,
\end{equation}
where $P$ is simply the product of the various priors for the
parameters, $P=\prod_{i=1}^{d}P_i$.

In the following we describe the principal steps of our algorithm:
\begin{itemize}
\item
We do a first sampling of the likelihood function with $N_{COV}$
 sampling points $\vec{\theta}$ and
we calculate the covariance matrix
$C=\langle\,(\vec{\theta}-\bar{\vec{\theta}})(\vec{\theta}-\bar{\vec{\theta}})^T\rangle$,
which is symmetric and positive definite
\item
A square matrix which is symmetric and positive definite can be
written as the product of a lower triangular and an upper triangular
matrices (Cholesky decomposition):
\begin{equation}
C=Q\cdot\,Q^T.
\end{equation}
We use the CHOLDC routine (see\citet{NR}, par.$2.9$) to calculate the matrix
$Q$. In general we can write the likelihood function as:
\begin{equation}
L(\vec{\theta})\propto\,\exp\left(-\frac{1}{2}\vec{\theta^T}[C^{-1}]\vec{\theta}\right),
\end{equation}
where $C$ is the covariance matrix; if $\vec{y}=Q^{-1}\vec{\theta}$ we now have:
\begin{equation}
L(\vec{y})\propto\,\exp\left(-\frac{1}{2}\vec{y}^T\vec{y}\right)
\end{equation}
\item
We now choose to sample $\vec{y}$ rather than $\vec{\theta}$, so our
sampling should be very efficient. In fact, most sampling points
will be generated in the subspace of the parameter space where the
likelihood function is not zero.

It is clear that, changing our variables, the integral of the
Likelihood will be given by:
\begin{equation}
\int_D\,L(\vec{\theta})d\vec{\theta}=\int_{D'}L(\vec{y})|\det(Q^{-1})|d\vec{y}.
\end{equation}
We perform $m$ statistically independent evaluations of the new
function using $N$ sampling points for each iteration. The iterations
are independent but they do assist each other because the algorithm
uses each one to improve the sampling grid for the next one. The
results of $m$ iterations are combined into a single best answer and
its estimated error, by standard inverse variance weighting.  We also compute $\chi^2$
to check that the best-fitting solutions are acceptable statistical fits.
%
\end{itemize}

Our results are clearly dependent on two parameters, the number $m$ of
iterations and the number $N$ of sampling points for each iteration. The more
iterations or sampling points are used, the more accuracy is reached.
In Figures \ref{fig1}-\ref{fig2} we show how much the standard
deviation depends on $m$ and $N$.
\begin{figure}
\begin{center}
\label{fig1}
\includegraphics[width=8cm]{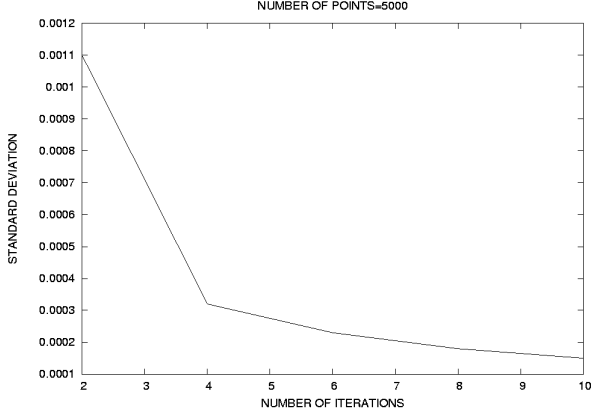}
\end{center}
\caption{Standard deviation in function of the number of iterations
  $m$ used (with $N=5000$ for each iteration) in the calculation of the
Likelihood function with $3$ parameters ($\Omega_m, w, H_0$). The
  value of the integral is: $I=0.04915$}
\label{fig1}
\end{figure}
\begin{figure}
\begin{center}
\label{fig2}
\includegraphics[width=8cm]{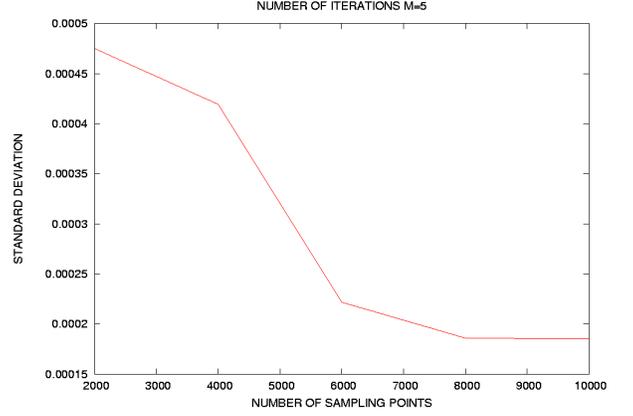}
\end{center}
\caption{The same of Figure \ref{fig1} but the standard deviation is plotted
  in function of the number N of sampling points (with $m=5$)}
\label{fig2}
\end{figure}
\begin{figure}
\begin{center}
\includegraphics[width=8cm]{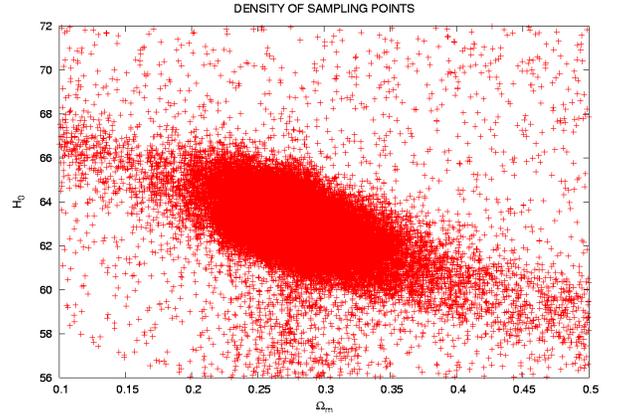}
\end{center}
\caption{Density of sampling points in the $2$-dimensional space
  $\Omega_m-H_0$}
\label{fig3}
\end{figure}
In Figure \ref{fig3} we can see the density of sampling points in
the $\Omega_m-H_0$ plane; the density will be greater in the
subspace where the Likelihood function lives and we can also see the
principal directions of the function. In general, implementing the
routine for the calculation of the covariance matrix, the relative
error $\frac{\sigma_I}{I}$ in the calculation of the integral is
lowered by a factor $10$, for the same number of function
evaluations. In our analysis we always use $m=5$ iterations and from
$N=5000$ to $30000$ sampling points (depending on the dimension of
the parameter space and on the size of the prior space), except for
MODEL V for which $N=100000$.
We reach an uncertainty of $\sim10^{-3}$ in
$\Delta\,\ln(E)$. We also used $N_{COV}=20000$ in the first
iteration for calculating the covariance matrix: this large number
is justified by the fact that it's important to achieve a good
estimate of the covariance matrix, to have good ``principal
directions'' for the next samplings, especially when we handle with
several parameters.

\section{Results}

Let us first analyze the full \citet{riess2007} data in combination with CMB and BAO.
The main results of this analysis are reported in Table 1 and in
Figures $4$-$8$.  The best-fitting parameter values are the means obtained from the
full posterior probability distribution, rather than the maximum likelihood values.  The
standard deviations are similarly obtained from integration over the
posterior.

\begin{table*}
\centering
 \begin{minipage}{140mm}
\caption{The parameter constraints together with the mean value of
  $\Delta\,\ln(E)$ and the minimum chisquared for the six models
  considered using the $182$ supernovae by Riess et al. 2007
  \citet{riess2007} binned with $\Delta\,z=0.05$ up to $z=1.7$ (about $34$
bins). The (unnormalized) $\ln(E)$ for the $\Lambda$CDM model is
 $\ln(E)=-16.2466\pm0.0002$. The uncertainty in the value of
  $\Delta\,\ln(E)$ is calculated making use of the usual formula for
  the propagation of the uncertainty of a variable
  $x=x(\vec{y})\rightarrow\sigma_x=
\sqrt{\sum_{i=1}^N\big(\frac{\partial\,x}{
\partial\,y_i}\big)^2\sigma_{x_i}^2}$.
 In our case $\Delta\,\ln(E)=\ln(\frac{E_1}{E_0})\rightarrow
\sigma_{\Delta\,\ln(E)}=\sqrt{\big(\frac{\sigma_{E_0}}{E_0}\big)^2+\big(\frac{\sigma_{E_1}}{E_1}\big)^2}$.}

\begin{tabular}{||p{4cm}||*{3}{c|}|}
\hline
\hline
\bfseries  Constraints  &  $\Delta\,lnE$     &  $\chi^2_{Min}$ & Model  \\
\hline
$\Omega_m=0.28\pm0.03$  &   0.0     &    24.39       &       I  \\
$H_0=64.5\pm0.09$       &                     &                &          \\
\hline
\hline
$\Omega_m=0.27\pm0.03$   &  $-0.222\pm0.005$   &    22.43    &      II   \\
$H_0=63.4\pm1.1$ &                     &           &          \\
$w<-0.84$ at $1\sigma$ &                       &           &           \\
$w<-0.73$ at $2\sigma$&          &         &  \\
\hline
\hline
$\Omega_m=0.27\pm0.03$    &  $-1.027\pm0.002$   &    22.43    &      III    \\
$H_0=63.4\pm1.1$          &                     &             &            \\
$w=-0.86\pm0.1$ &                     &             &             \\
\hline
\hline
$\Omega_m=0.28\pm0.04$    &  $-1.118\pm0.015$   &    21.47    &      IV    \\
$H_0=63.8\pm1.4$  &                     &             &            \\
$w_0=-1.03\pm0.25$&                    &             &            \\
$w_a=0.76^{+--}_{-0.91}$         &     &             &            \\
\hline
\hline
$\Omega_m=0.27\pm0.03$   &  $-1.059\pm0.008$   &    21.38     &    V    \\
$H_0=63.5\pm1.1$  &                     &              &          \\
$w_0=-0.85\pm0.12$&                   &              &           \\
$w_1=-0.81\pm0.21$&                   &              &           \\
$a_s$ unconstrained &                     &              &         \\
$q$ unconstrained     &                  &              &          \\
\hline
\hline
$\Omega_m=0.30\pm0.05$    &  $-1.834\pm0.006$    &    21.52     &   VI    \\
$H_0=63.5^{+1.8}_{-1.2}$ &                      &              &          \\
$w_0=-1.08_{-0.30}^{+0.24}$&                     &              &         \\
$w_1=0.78_{-0.57}^{+0.83}$&                      &              &        \\
\hline
\hline
\end{tabular}
\end{minipage}
\end{table*}

\vskip1cm

\begin{figure}
\begin{center}
\includegraphics[width=9cm]{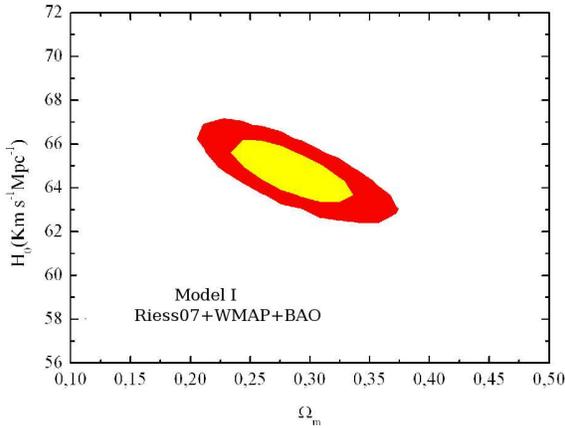}
\end{center}
\caption{Likelihood contours at $68\%$ and $95\%$ (two-parameter)
for MODEL I} \label{fig4}
\end{figure}

\begin{figure}
\begin{center}
\includegraphics[width=9cm]{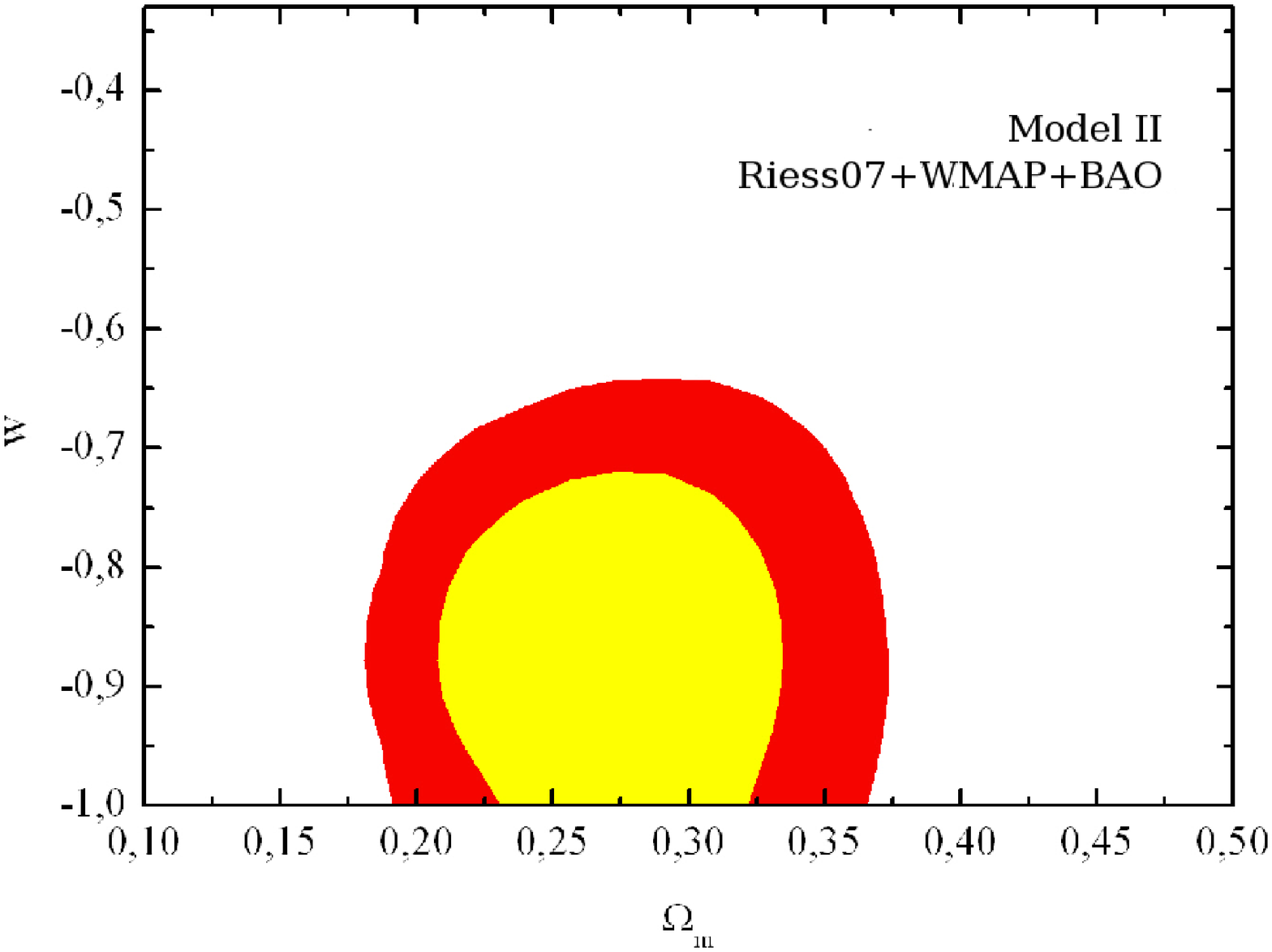}
\end{center}
\caption{Likelihood contours at $68\%$ and $95\%$ for MODEL II}
\label{fig5}
\end{figure}

\begin{figure}
\begin{center}
\includegraphics[width=9cm]{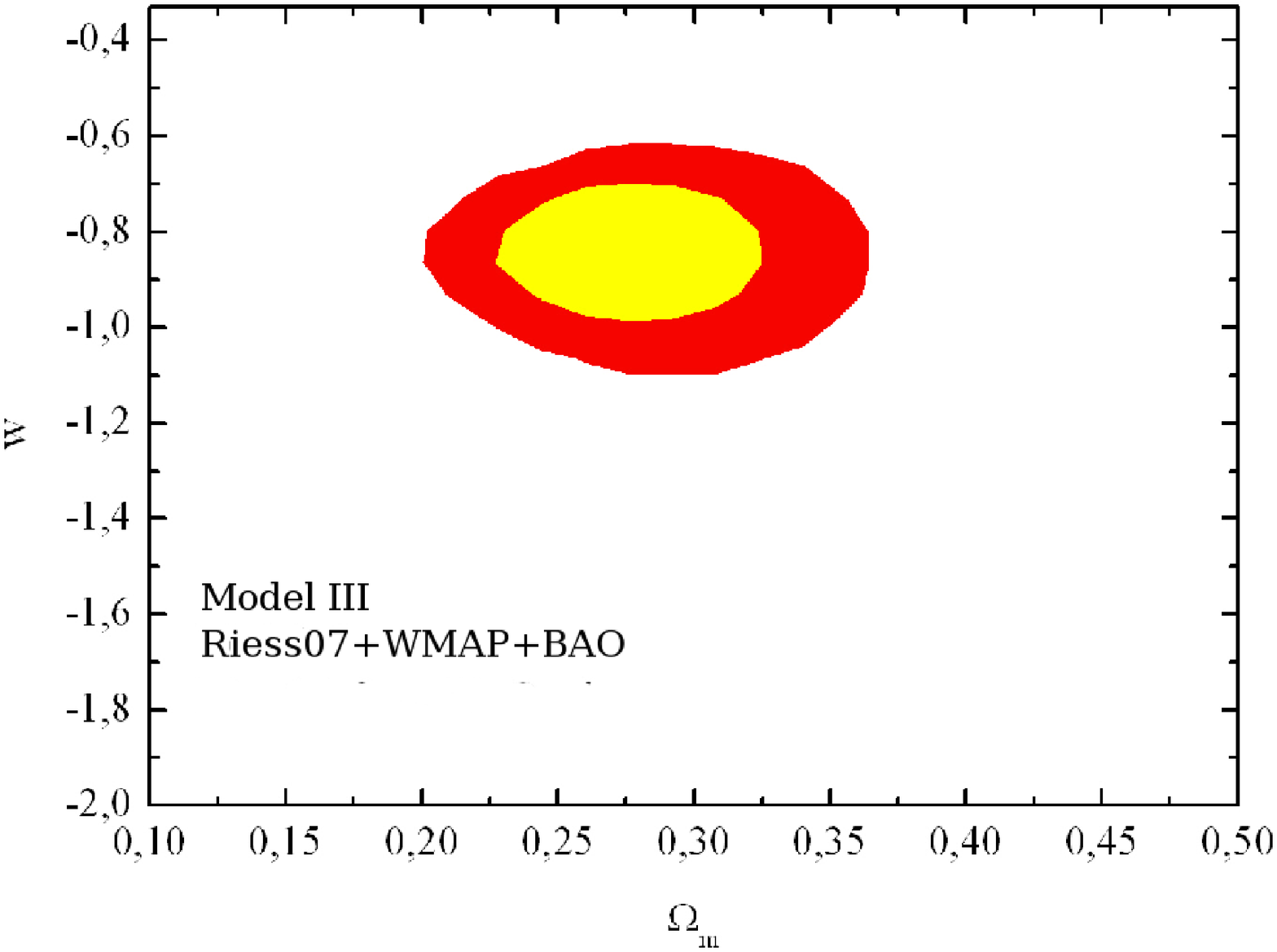}
\end{center}
\caption{Likelihood contours at $68\%$ and $95\%$ for MODEL III}
\label{fig6}
\end{figure}

\begin{figure}
\begin{center}
\includegraphics[width=9cm]{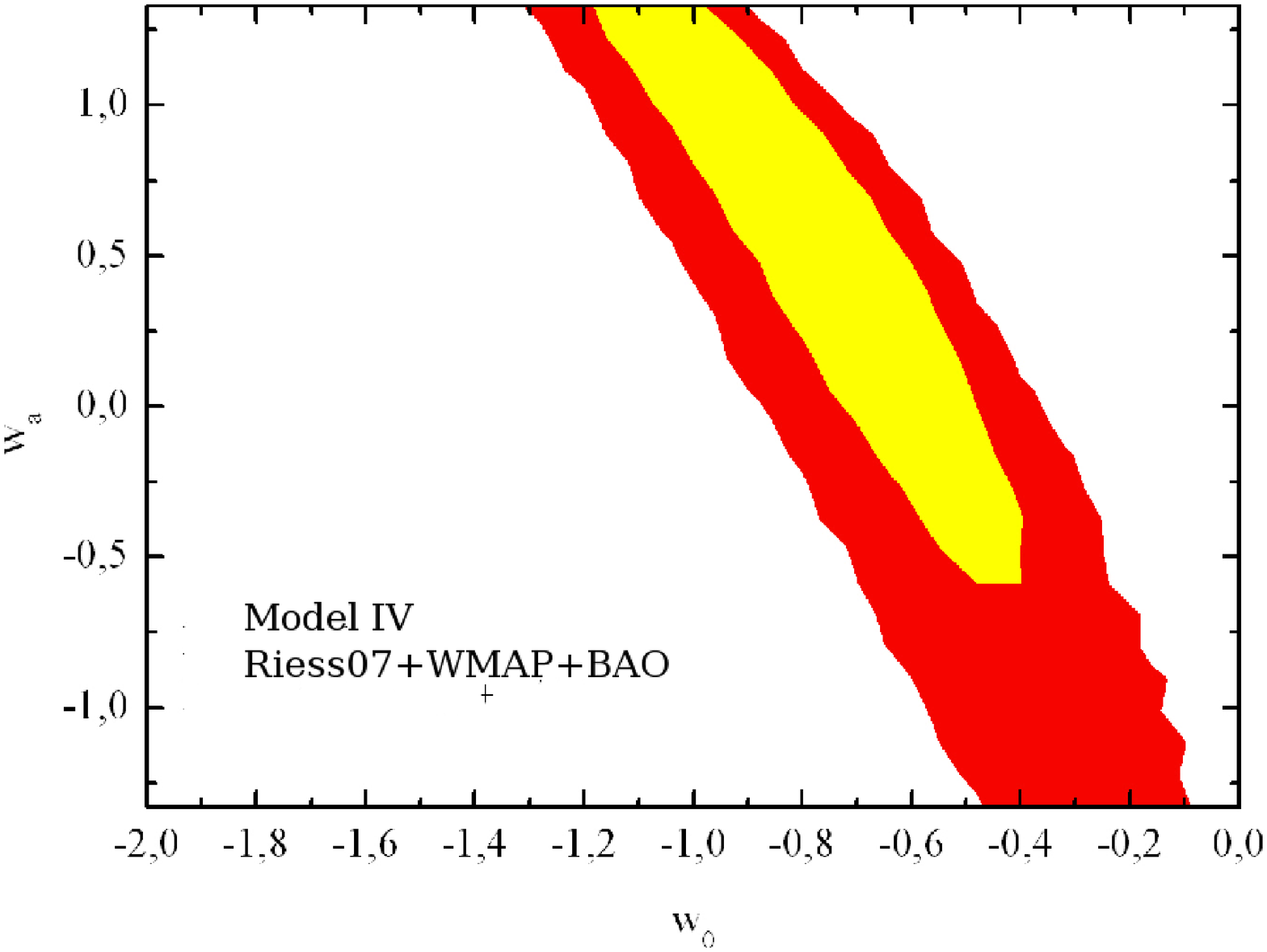}
\end{center}
\caption{Likelihood contours at $68\%$ and $95\%$ for MODEL IV}
\label{fig7}
\end{figure}


\begin{figure}
\begin{center}
\includegraphics[width=9cm]{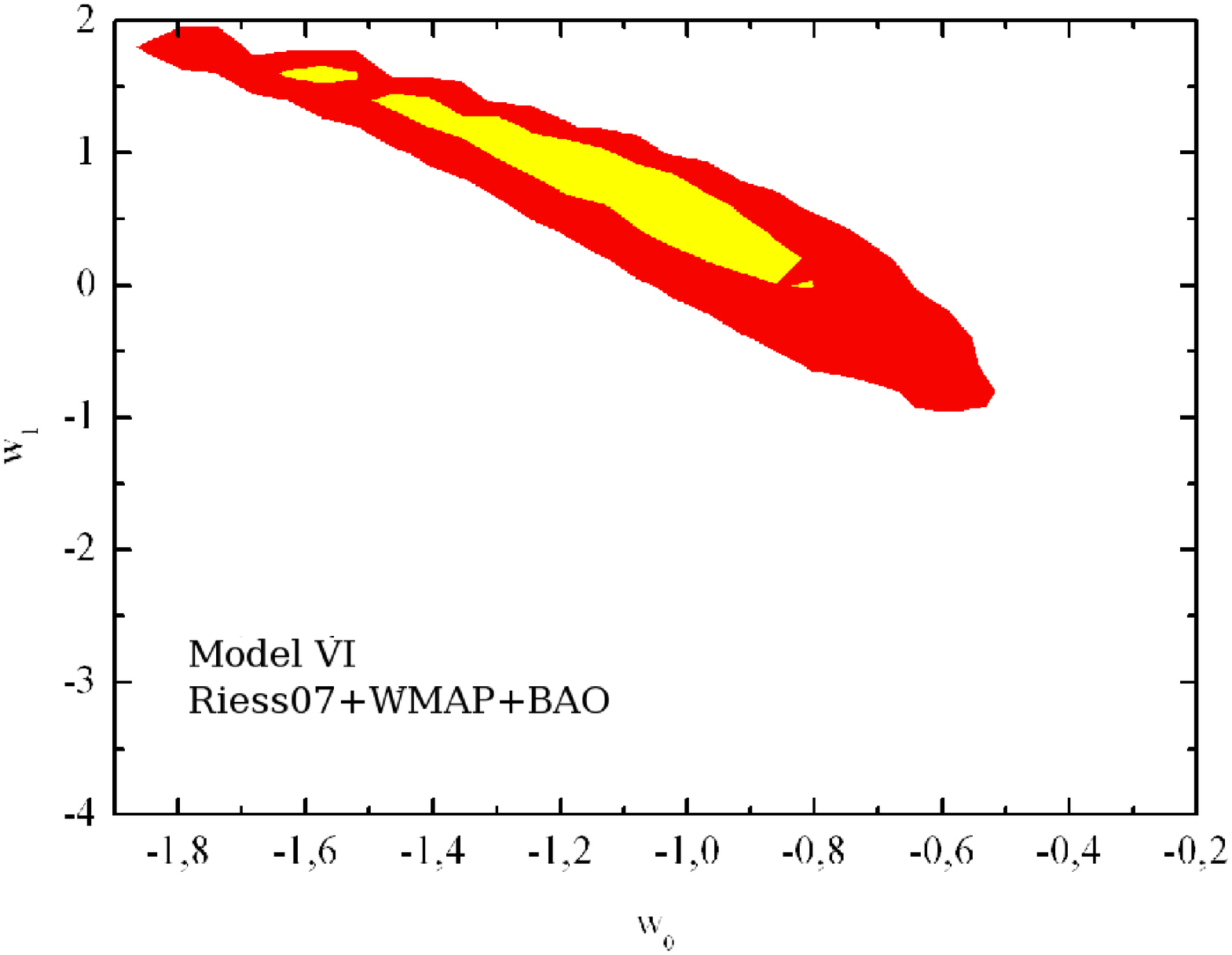}
\end{center}
\caption{Likelihood contours at $68\%$ and $95\%$ for MODEL VI}
\label{fig9}
\end{figure}

As we can see a cosmological constant is preferred by the data and is
always compatible with it, independently of the model considered. The
constraints we obtain on the equation of state parameter, assumed as a
constant, are $w< -0.84$ in the case of $w \ge -1$ and $w=-0.86\pm0.1$
at $68 \%$ c.l.  when models with $w<-1$ are included. Those
constraints are compatible and of the same order of magnitude of the
previous constraints reported by \citet{liddle3} but where the new
SN-Ia dataset of \citet{riess2007} was not considered. However the
evidence for the $w<-1$ case is worse by $\Delta \ln (E) \sim 1$,
i.e. there is no indication from the data that we should extend the
parameter space to these phantom dark energy models.

The same happens when we consider models with an equation of state
parameter which varies with redshift.

For the models I-IV, which have the best Evidences and the same
parametrization (with
$w_0=-1$ and $w_a=0$ for MODEL I and $w_a=0$ for MODEL II-III)
we have also considered results coming from the
\emph{Bayesian model averaging}; as explained in \citet{liddle3},
because of our ignorance about the true cosmological model, we may
think that the probability distribution of the parameters is a
superposition of its distributions in different models, weighted by
the relative model probability, as in quantum mechanics, where the
state of a physical system is a superposition of its possibilities
until a measurement determines the collapse in a single eigenstate.

If we convert the $\Delta\,\ln(E)$ into posterior probabilities,
assuming equal prior probabilities, we have $40.2\%$, $32.2\%$,
$14.4\%$, $13.2\%$ for models I-II-III-IV.

The constraints on the cosmological parameters from the Bayesian model
averaging of models I-II-III-IV are: $\Omega_m=0.27\pm0.03$,
$H_0=63.2^{+1.8}_{-1.2}$, $w_0=-1.0\pm0.1$
at $1\sigma$, $w_a=0.0^{+0.02}_{-0.03}$ at $2\sigma$. The
confidence limits in $w_a$ are exacly zero at $1\sigma$; this is
because the probability distribution for this parameter is a delta
function for models I-II-III and it is superimposed to the extended
tails of model IV (as explained in \citet{liddle3}).

It is also interesting to consider the effects on the cosmological
parameters if we remove from the dataset the supernovae with $z>1$.
In Table 2 we report our principal results.

\begin{table*}

\centering
\begin{minipage}{140mm}

\caption{The parameter constraints together with the mean value of
  $\Delta\,\ln(E)$ and the minimum chisquared for the six models
  considered when $z<1$, using the $166$ supernovae with $z<1$ of
  Riess et al. $2007$ binned with $\Delta z=0.05$ (about $20$ bins).}

\begin{tabular}{||p{4cm}||*{3}{c|}|}
\hline
\hline
\bfseries  Constraints  &  $\Delta\,lnE$      & $\chi^2_{Min}$ & Model  \\
\hline
$\Omega_m=0.27\pm0.04$  &   0.0     &    16.43       &       I  \\
$H_0=64.6\pm0.09       $&                     &                &          \\
\hline
\hline
$\Omega_m=0.27\pm0.04$   &  $-0.223\pm0.004$   &    14.54    &      II   \\
$H_0=63.4\pm1.3$ &                     &           &          \\
$w<-0.82$ at $1\sigma$ &                       &           &           \\
$w<-0.73$ at $2\sigma$&          &         &  \\
\hline
\hline
$\Omega_m=0.27\pm0.04$    &  $-1.022\pm0.004$   &    14.54    &      III    \\
$H_0=63.4\pm1.2$          &                     &             &            \\
$w=-0.86\pm0.1$ &                     &             &             \\
\hline
\hline
$\Omega_m=0.29\pm0.04$    &  $-1.090\pm0.011$   &    13.33    &      IV    \\
$H_0=63.6\pm1.2$  &                     &             &            \\
$w_0=-0.98^{+0.26}_{-0.23}$&                    &             &            \\
$w_a=0.72^{+--}_{-1.08}$ &     &             &            \\
\hline
\hline
$\Omega_m=0.28\pm0.04$   &  $-1.020\pm0.008$   &    13.00     &    V    \\
$H_0=63.4\pm1.0$  &                     &              &          \\
$w_0=-0.88\pm0.12$&                      &              &           \\
$w_1=-0.63^{+0.2}_{-0.27}$&                   &              &           \\
$a_s=$ (unconstrained)&               &              &         \\
$q=$ (unconstrained)&                &              &          \\
\hline
\hline
$\Omega_m=0.30\pm0.05$    &  $-1.691\pm0.01$    &    12.95     &   VI    \\
$H_0=64.3\pm1.2$  &                      &              &          \\
$w_0=-1.17\pm0.37$&                     &              &         \\
$w_1=$ (unconstrained)&                      &              &        \\
\hline
\hline
\end{tabular}
\end{minipage}
\end{table*}

As we can see, there are no significant differencies on the mean
values however the error bars are generally reduced by a $\sim 30 \%$.
Models with varying-with redshift equation of state have
a slightly better evidence but with
a cosmological constant is still favoured.

It's also useful to check if our results are the same when we
consider a different dataset. To this extent, we use the $57$
supernovae coming from the ESSENCE Supernova Suvey \citet{wood} in
combination with the $38$ nearby supernovae (with $0.023<z<0.15$) of
Riess et al. 2007 (\citet{riess2007}). We do not consider models with
evolving dark energy, because in this analysis we limit our redshift
range to $z<0.670$.

\begin{table*}
 \centering
 \begin{minipage}{140mm}
\caption{Parameter constraints and Evidence for MODEL I-II-III, using
  the 95 supernovae of ESSENCE+nearby \citet{riess2007} binned with
 $\Delta\,z=0.05$ up to $z=0.65$ ($15$ bins)}.
\begin{tabular}{||p{4cm}||*{3}{c|}|}
\hline
\hline
\bfseries  Constraints  &  $\Delta\,lnE$      & $\chi^2_{Min}$ & Model  \\
\hline
$\Omega_m=0.25\pm0.04$  &   0.0     &    10.73       &       I  \\
$H_0=64.7\pm0.09       $&                     &                &          \\
\hline
\hline
$\Omega_m=0.27\pm0.04$   &  $-0.258\pm0.004$   &    9.28    &      II   \\
$H_0=63.6\pm1.3$ &                     &           &          \\
$w<-0.80$ at $1\sigma$ &                       &           &           \\
$w<-0.67$ at $2\sigma$&          &         &  \\
\hline
\hline
$\Omega_m=0.27\pm0.04$    &  $-1.027\pm0.003$   &    9.28    &      III    \\
$H_0=63.5\pm1.2$          &                     &             &            \\
$w=-0.86\pm0.11$ &                     &             &             \\
\hline
\hline
\end{tabular}
\end{minipage}
\end{table*}

The results are reported in Table 3. The results are fully compatible
with those from the previous analysis and, again, they provide a
substantial evidence for a constant $w$, with a cosmological constant
being preferred.

\section{Conclusions}

In this paper we have carried out a Bayesian model selection analysis
of several dark energy models using the new data of high redshift
supernovae of Riess et al. 2007 (\citet{riess2007})
and from the ESSENCE survey (\citet{wood}), together with Baryonic Acoustic
Oscillations and Cosmic Microwave Background Anisotropies. To this
extent, we have developed a new algorithm to calculate the Bayesian
Evidence which is fast (less than $1$ hour for each calculation of
the Evidence on a $2 GHz$ CPU) and very accurate (a relative uncertainty
$\frac{\sigma_{\Delta\,\ln(E)}}{\Delta_{\ln(E)}}<10^{-2}$ in $10^5$
likelihood evaluations).

We find that with current observational data the usual $\Lambda$CDM
model is slightly preferred with respect to dark energy models with
equations of state in the range $-1\leq\,w\leq0$ and substantially
preferred to dark energy models with $-2\leq\,w\leq0$ or to dark
energy models with an equation of state which evolves with time.

However we would like also to stress that it may be premature to
reject models only on the basis of Bayesian model selection; in
general, the simplest models may not be the ``true'' model. In this
way, until we have a theoretical explanation of the accelerated
expansion of the Universe, one should keep an open mind to all the
alternatives to the $\Lambda$CDM scenario, even if, at the moment,
it seems the simplest description of our Universe.

\section{Acknowledgments}

Paolo Serra thanks University of Edinburgh for support during his
visit and members of the Royal Observatory of Edinburgh for their
kind ospitality.  We thank Andrew Liddle for useful comments.


\end{document}